\documentclass[11pt,epsf]{article}

\usepackage{epsfig}
\usepackage{amssymb}
\usepackage{graphicx}
\usepackage{color}
\usepackage{subfigure}

\begin{document}

\baselineskip 6mm
\renewcommand{\thefootnote}{\fnsymbol{footnote}}


\newcommand{\nc}{\newcommand}
\newcommand{\rnc}{\renewcommand}



\newcommand{\tcb}{\textcolor{blue}}
\newcommand{\tcr}{\textcolor{red}}
\newcommand{\tcg}{\textcolor{green}}


\def\be{\begin{equation}}
\def\ee{\end{equation}}
\def\ba{\begin{array}}
\def\ea{\end{array}}
\def\bea{\begin{eqnarray}}
\def\eea{\end{eqnarray}}
\def\nn{\nonumber\\}


\def\ct{\cite}
\def\la{\label}
\def\eq#1{(\ref{#1})}


\def\a{\alpha}
\def\b{\beta}
\def\g{\gamma}
\def\G{\Gamma}
\def\d{\delta}
\def\D{\Delta}
\def\e{\epsilon}
\def\et{\eta}
\def\ph{\phi}
\def\Ph{\Phi}
\def\ps{\psi}
\def\Ps{\Psi}
\def\k{\kappa}
\def\l{\lambda}
\def\L{\Lambda}
\def\m{\mu}
\def\n{\nu}
\def\th{\theta}
\def\Th{\Theta}
\def\r{\rho}
\def\s{\sigma}
\def\S{\Sigma}
\def\ta{\tau}
\def\o{\omega}
\def\O{\Omega}
\def\pr{\prime}
\def\z{\zeta}


\def\half{\frac{1}{2}}

\def\goto{\rightarrow}

\def\na{\nabla}
\def\grad{\nabla}
\def\curl{\nabla\times}
\def\div{\nabla\cdot}
\def\pa{\partial}
\def\fr{\frac}
\def\sq{\sqrt}
\def\inf{\infty}

\def\bra{\left\langle}
\def\ket{\right\rangle}
\def\lb{\left[}
\def\lc{\left\{}
\def\ls{\left(}
\def\lp{\left.}
\def\rp{\right.}
\def\rb{\right]}
\def\rc{\right\}}
\def\rs{\right)}
\def\fr{\frac}

\def\vac#1{\mid #1 \rangle}


\def\td#1{\tilde{#1}}
\def\check{ \maltese {\bf Check!}}


\def\Tr{{\rm Tr}\,}
\def\det{{\rm det}}


\def\bc#1{\nnindent {\bf $\bullet$ #1} \\ }
\def\ch {$<Check!>$ }
\def\ss {\vspace{1.5cm}}

\begin{titlepage}

\hfill\parbox{5cm} { }

\vspace{25mm}


\vspace{2cm}

\begin{center}
{\Large \bf Holographic aspects of a relativistic non-conformal theory \\
}

\vskip 1. cm
  {Chanyong Park$^a$\footnote{e-mail : cyong21@sogang.ac.kr}}

\vskip 0.5cm

{\it $^a\,$Center for Quantum Spacetime (CQUeST), Sogang University, Seoul 121-742, Korea}\\

\end{center}

\thispagestyle{empty}

\vskip2cm


\centerline{\bf ABSTRACT} \vskip 4mm

\vspace{1cm}
We study a general $D$-dimensional Schwarzschild-type black brane solution of the 
Einstein-dilaton theory and derive, by using the holographic renormalization, its thermodynamics 
consistent with the geometric results.
Using the membrane paradigm,
we calculate the several hydrodynamic transport coefficients and compare them
with the results obtained by the Kubo formula, which
shows the self-consistency of the gauge/gravity duality in the relativistic non-conformal
theory. In order to understand more about the relativistic non-conformal theory, we
further investigate the binding energy, drag force and holographic entanglement 
entropy of the relativistic non-conformal theory. 

\vspace{2cm}

\end{titlepage}

\renewcommand{\thefootnote}{\arabic{footnote}}
\setcounter{footnote}{0}


\section{Introduction}

For the last decade, the AdS/CFT correspondence  \cite{maldacena} 
has been one of the interesting subjects for understanding the strongly interacting quantum field theory (QFT). 
Applying the AdS/CFT 
correspondence to QCD or the condensed matter system provided many interesting physical
results like the phase structures \cite{Erlich:2005qh,Salvio:2012at}, 
the universal ratio between the shear viscosity and entropy 
\cite{Policastro:2002se}-\cite{Springer:2008js} , 
holographic superconductor \cite{Hartnoll:2009sz}-\cite{Gauntlett:2009dn} and strange metallic behavior \cite{Hartnoll:2009ns}, etc.
Recently, after assuming the gauge/gravity duality it was shown that 
the dual field theory of the Einstein-dilaton gravity can be described by a relativistic
non-conformal theory \cite{Kulkarni:2012re}. In this model, due to the running dilaton, 
the DC conductivity obtained by the Kubo formula \cite{Policastro:2002se,Policastro:2002tn,Kulkarni:2012re} shows the unexpected behavior, which may describe electrolyte
or some chemical compounds. 
The real physics is not usually conformal except some critical phenomena like the phase transition and the RG fixed points. Therefore, it is required to generalize the AdS/CFT correspondence to the non-conformal case. Here, we simply call such a generalized correspondence the gauge/gravity duality.
Actually, it is not hopeful to prove the gauge/gravity duality because even in the AdS space
there is no direct proof of the AdS/CFT correspondence. Instead, we will try to find some evidences for the gauge/gravity duality of a non-conformal theory.

In the Einstein-dilaton theory with a Liouville potential, there exists a Schwarzschild-type
black brane solution which we call an Einstein-dilaton black brane (EdBB). Since its 
asymptotic geometry is not the AdS space and the induced metric on the boundary 
is given by the Minkowski metric, the gauge/gravity duality says that the dual theory should be 
a relativistic non-conformal theory. 
In order to understand more physical properties, one can apply the holographic renormalization
to the Einstein-dilaton theory. In this paper, after finding an appropriate counter term 
we show that the resulting on-shell action and the boundary stress tensor are finite.
Furthermore, we check the self-consistency of the gauge/gravity duality by showing that the thermodynamic quantities derived from the boundary 
stress tensor coincide with the results of the EdBB geometry.

Another interesting issue related to the self-consistency is the hydrodynamics. 
In the linear response theory of the QFT
the macroscopic properties can be determined by the transport coefficients.
Moreover, they can be represented by the background thermal quantities. For example, 
consider a thermal system with an energy density $\e$ and pressure $P$. 
Then, the momentum diffusion constant of this system is given by (see \cite{Hartnoll:2009sz}
and references therein)
\be		\la{res:QFT}
{\cal D}_s = \fr{\et_s}{\e + P} ,
\ee
where $\et_s$ is the shear viscosity. Note that in this calculation the microscopic details
of the system is not important.
If the EdBB is really dual to a relativistic non-conformal theory, the dual system should
also satisfy this relation. 
Using the universality of the ratio between the shear viscosity and the entropy
density \cite{Iqbal:2008by,Hartnoll:2009sz}
\be			\la{des:univers}
\frac{\eta_s}{s} = \frac{1}{4 \pi} ,
\ee  
as well as the thermodynamic results of the Einstein-dilaton theory, the momentum diffusion constant becomes
\be		\la{res:mdcthermo}
{\cal D}_s (\infty) = \frac{1}{4 \pi T_H} .
\ee
An alternative way to obtain the momentum diffusion constant is to investigate the 
holographic hydrodynamics of the metric fluctuations. 
For the self-consistency
those two results should be the same. We show by applying the membrane paradigm
\cite{Iqbal:2008by}
that the momentum diffusion constant obtained by the holographic method
really satisfies the above relation \eq{res:mdcthermo}.

In order to understand more physical features of the relativistic non-conformal theory, 
we further investigate the holographic binding energies probed by an F1- and D1-string. 
In the boundary theory point of view, an F1- or D1-string represents the bound state of a pair of particles (fundamental excitations) or monopoles (solitons) respectively. 
In the AdS space, there is no physical difference between an F1- and D1-string 
due to the conformality or the trivial dilaton profile. However, in the relativistic non-conformal theory a monopole is distinguished from a particle because of the non-trivial coupling constant.
The holographic results show that the binding energies of particles and monopoles 
are stronger in the non-conformal theory than in the conformal one.
We also investigate the drag forces of a particle and monopole in the non-conformal medium.
For a relativistic particle and monopole, the momentum exponentially decreases as
time evolves. The dissipation rate of the momentum is proportional to temperature
with a positive power depending on the non-conformality. In the non-relativistic case,
the momentum decays with an inverse power law for a particle and with a power law 
for a monopole, in which the dissipation power is again determined by the non-conformality.
Finally, we investigate the holographic entanglement entropy of the relativistic non-conformal 
theory. Recently, there was an interesting conjecture that in a small subsystem 
the entanglement temperature (or `effective temperature') has a universal feature 
proportional to the inverse of the size \cite{Bhattacharya:2012mi}. 
Such a universality of the entanglement temperature also appears in the
holographic relativistic non-conformal theory.

The rest of the paper is organized as follows: In Sec. 2, we explain our conventions
and summarize the black brane thermodynamics. 
For checking the gauge/gravity duality of the relativistic non-conformal theory,
in Sec. 3, we rederive the same thermodynamics from the boundary energy-momentum tensor
constructed by the holographic renormalization. Applying the membrane paradigm in Sec. 4, we
also show that the momentum diffusion constant obtained in the EdBB gives rise to
the consistent result with the EdBB thermodynamic and satisfies the QFT relation \eq{res:QFT}.
Based on these self-consistencies of the gauge/gravity duality in the relativistic non-conformal
theory, we further investigate the binding energies of particles and monopoles in Sec. 5
and the drag forces in Sec. 6.
In Sec. 7, by calculating the holographic entanglement entropy,
we show that the universal feature of the entanglement temperature conjectured in \cite{Bhattacharya:2012mi} is still valid even in the relativistic non-conformal theory.
Finally, we finish our work with some concluding remarks in Sec. 8.


\section{$D$-dimensional Einstein-dilaton black brane}

Let us consider a $D$-dimensional Einstein-dilaton theory with a Liouville potential
in a Lorentzian signature 
\be
S_{Ed} = \fr{1}{16 \pi G} \int_{\cal M} d^{D} x \sq{-g} \lb {\cal R} - 2 (\pa \phi)^2 
- 2 \L e^{\eta \phi}  \rb ,
\ee
where $\L$ is a negative constant. We simply call $\L$ a cosmological constant because it 
really becomes a cosmological constant for $\et = 0$. In the above, $\et$ is an arbitrary
constant representing the non-conformality of the dual theory.
Taking a logarithmic profile for the dilaton field 
\be			\la{ans:dilatonpro}
\ph(r) = \ph_0 - k_0 \log r ,
\ee
where $\ph_0$ and $k_0$ are two integration constants, 
an effective cosmological constant $\L_{eff} \equiv \L e^{\et \ph_0}$ 
without loss of generality can be set to be 
\be			\la{res:effcosmo}
\L_{eff} = - \frac{4 (D-2) \lb 8 (D-1) - (D-2) \eta^2 \rb }{ \lb 8 + (D-2) \eta^2 \rb^2} .
\ee
Note that since $\L$ is negative $\L_{eff}$ should also be negative. This fact implies that
$\et$ is below the Gubser bound, $\eta^2 < 8 (D-1) / (D-2)$) \cite{Gubser:2000nd,Gursoy:2007cb,Gursoy:2007er,Gouteraux:2011ce}. 
Consequently, $\ph(r)$ is refined to 
\be
\ph(r) = - k_0 \log r .
\ee
In terms of the refined dilaton field and the effective cosmological constant,  
the Einstein equation and the equation of motion for a dilaton become
\begin{eqnarray}
{\cal R}_{\mu\nu}-\frac{1}{2} {\cal R} g_{\mu\nu}+ g_{\mu\nu} \ \L_{eff} \ e^{\et \ph} &=& 2 \partial_{\mu}\phi
\partial_{\nu}\phi-  g_{\mu\nu}(\pa \phi)^{2} ,  \la{eq:Einstein}\\
\frac{1}{\sqrt{-g}} \partial_{\mu}(\sqrt{-g} g^{\m \n} \partial_{\n}\phi) &=& 
\frac{1}{2} \L_{eff} \ \et \ e^{\et \ph} \la{eq:scalar}.
\end{eqnarray}
Together with the logarithmic dilaton profile, the EdBB metric satisfying
above two equations is given by
\bea	\la{eq:asymp}
ds^{2} = - r^{2 a_1} f(r) dt^{2}+\frac{dr^{2}}{r^{2 a_1} f(r)}+r^{2 a_1} \d_{ij} dx^i dx^j  ,
\eea
with a black brane factor 
\be
f(r) = 1 - \frac{r_h^c}{ r^{c}} ,
\ee
where $i$ and $j$ represent the spatial directions of the boundary space and the other 
parameters are
\bea	\la{sol:fivepara}
k_0 &=& \frac{2 (D-2) \eta}{8+ (D-2) \eta^2} , \nn
a_1  &=&  \frac{8}{8 + (D-2) \eta^2}  , \nn
c &=& \frac{8 (D-1) - (D-2) \eta^2}{8 + (D-2) \eta^2} .
\eea

The asymptote of the EdBB metric reduces to
\be
ds^2 = \frac{1}{r^{2 a_1} } dr^2 + r^{2 a_1} \ls - dt^2 +\d_{ij} dx^i dx^j  \rs ,
\ee
where the Poincare symmetry $ISO(1,D-2)$ of the boundary hypersurface at a fixed $r$ is
manifest. This fact implies that the dual field theory should be described by a relativistic QFT. 
For $\eta=0$, the background geometry reduces to the asymptotic AdS space
with an effective cosmological constant $\L_{eff}$ 
\be
\L_{eff} = - \frac{(D-1) (D-2)}{2} ,
\ee
which is exactly that of a general $D$-dimensional AdS space with an unit AdS radius $R=1$.
For $\et \ne 0$, the asymptotic geometry is not the AdS space anymore and instead reduces
to the hyperscaling violation form \cite{Goldstein:2009cv}-\cite{Gouteraux:2011qh}.
 Therefore, 
one can easily see that the dual 
field theory is not conformal.

Before concluding this section, let us summarize thermodynamics of the EdBB obtained from the metric \eq{eq:asymp}.
The Hawking temperature and the Bekenstein-Hawking entropy are 
\bea			\la{res:hawkingtemp}
T_H &=& \frac{1}{4 \pi} \frac{ 8  (D-1) + (D-2) \eta^2 }{8 + (D-2) \et^2} \ 
r_h^{\fr{8 - (D-2) \eta^2 }{8 + (D-2) \eta^2 }} ,	 \nn
S_{BH} 
&=& \frac{V_{D-2}}{4 G}  \ r_h^{\fr{8 (D-2)}{8 + (D-2) \et^2} } ,
\eea 
where $V_{D-2}$ means the spatial volume of the boundary space. 
Other thermodynamic quantities, the internal energy $E$ and the free energy $F$, are
given by
\bea			\la{res:thermodyEdbb}
E &=& \frac{ V_{D-2}}{8 \pi G} \frac{4 (D-2) }{ 8 + (D-2) \eta^2 }
\  r_h^{ \fr{ 8 (D-1) - (D-2) \eta^2}{ 8 + (D-2) \eta^2} },  \la{thres:internalenergy} \\
F &\equiv& E - T S_{BH} = - \frac{V_{D-2}}{16 \pi G} 
\frac{ 8 - (D-2) \eta^2  }{  8 + (D-2) \eta^2 } 
\ r_h^{\fr{8 (D-1) - (D-2) \eta^2}{ 8 + (D-2) \eta^2 }} . \la{thres:freeenergy} 
\eea
Using the pressure defined by $P = - \fr{\pa F}{\pa V_{D-2}}$, the thermodynamic quantities of the EdBB satisfy the first law of thermodynamics 
as well as the Gibbs-Duhem relation, $E + P V_{D-2} = T S_{BH}$.
Following the gauge/gravity duality, these thermodynamic quantities can also be reinterpreted
as those of the dual relativistic non-conformal theory.

\section{Holographic renormalization of the Einstein-dilaton theory}

In the AdS/CFT correspondence, the conformal symmetry usually
plays an important role to match spectra of gravity with their dual operators. In the
non-conformal case, although such a relation is not clear, 
we can still investigate some thermodynamic properties of the dual field theory through 
the gauge/gravity duality. If the on-shell gravity action 
is identified with the free energy of the dual field theory, we can easily 
derive the thermodynamic properties. In this section, we will show that the 
on-shell gravity action, after the appropriate holographic renormalization, really provides 
the consistent thermodynamics with the EdBB thermodynamics.

In order to describe a finite temperature system, it is more convenient to take into account 
an Euclidean version. With an Euclidean signature,
the gravitational action of the Einstein-dilaton theory can be rewritten as
\be			\la{act:gravity}
S_{gr} = S_{Ed} + S_{GH} ,
\ee 
with
\bea
S_{Ed} &=& - \fr{1}{16 \pi G} \int_{\cal M} d^{D} x \sq{g} \lb {\cal R} - 2 (\pa \phi)^2 
- 2 \L_{eff} \ e^{\eta \phi}  \rb , \nn
S_{GH} &=&  \fr{1}{8 \pi G} \int_{\pa {\cal M}} d^{D-1} x \sq{\g} \ \Th ,
\eea
where
$g_{\m \n}$ or $\g_{a b}$ is the Euclidean metric in the bulk or the induced
boundary metric respectively.
For a well-defined action variation, the Gibbons-Hawking
term $S_{GH}$ is required.
An extrinsic curvature tensor $\Th_{\m\n}$ is defined by
\be
\Th_{\m\n} = - \half \ls \na_{\m} n_{\n} + \na_{\n} n_{\m} \rs ,
\ee
where $\na_{\m}$ and $n_{\n}$ mean a covariant derivative and an unit
normal vector respectively. Since the Gibbons-Hawking term is a boundary term, 
it does not affect on the equations of motion. 

Before evaluating the on-shell gravity action, it is worth noting that  
the on-shell gravity action usually suffers from the divergence when the boundary is located at  
$r = \inf$. To remove such an UV divergence, we should add appropriate counter terms 
which make the on-shell gravity action become finite 
\cite{Balasubramanian:1999re,Batrachenko:2004fd,Son:2006em}. 
The correct counter term we find is
\be
S_{ct} = \fr{1}{8 \pi G}  \int_{\pa {\cal M}} d^{D-1} x \sq{\g} \ \ls \fr{8 (D-2) 
}{8 + (D-2) \et^2} \ e^{\et \ph /2} \rs .
\ee
For $\et = 0$ it reduces, after restoring the $AdS$ radius $R$, to the usual one for $AdS_D$ space
\cite{Balasubramanian:1999re}
\be
S_{AdS} = \fr{1}{8 \pi G}  \int_{\pa {\cal M}} d^{D-1} x \sq{ \g} \  \fr{(D-2) 
}{R} .
\ee
In general, there exist additional counter terms proportional to the boundary curvature scalar
or tensors. However, since the boundary space of the EdBB geometry is flat with a Poincare 
symmetry $SO(1,D-2)$, the counter terms associated with the curvature scalar or tensors 
automatically vanish. Therefore, the resulting renormalized action can be described by
\be
S = S_{Ed} + S_{GH} + S_{ct} .
\ee  
Since the on-shell gravity action reduces to a boundary term,  it
can be naturally interpreted as a boundary quantity. Following the strategy of the AdS/CFT
correspondence, it should be proportional to the free energy of the dual theory with providing the 
same thermodynamics derived in the EdBB geometry. Now, let us check that the renormalized
on-shell gravity action really reproduces the results of the EdBB in \eq{res:thermodyEdbb}.

First, consider the Einstein-dilaton action. Using the Einstein equation in \eq{eq:Einstein}, it
simply reduces to
\be			\la{act:Ed}
S_{Ed} = - \fr{1}{16 \pi G} \int_0^{\b} d \ta \int d^{D-2} x  \int_{r_h}^{r_0} dr 
\sq{g} \ \fr{4}{D-2} \ \L_{eff} \ e^{\et \ph} ,
\ee
where $\ta$ is an Euclidean time with a periodicity $\b$ and $\int d^{D-2} x  = V_{D-2}$ is 
the spatial volume of the boundary space. In the above, $r_0$ is introduced to denote the
position of the boundary which can be interpreted as an UV cutoff of the dual theory.
After inserting solutions, \eq{ans:dilatonpro} and \eq{res:effcosmo}, into \eq{act:Ed} and evaluating it, we finally reach to
\be
S_{Ed} = \fr{\b V_{D-2}}{16 \pi G} \fr{16}{8 + (D-2) \et^2} \ls r_0^c - r_h^c \rs ,
\ee
which diverges as $r_0 \to \inf$ because of $c>1$. 
Similarly, the Gibbons-Hawking term and the counter term 
result in 
\bea
S_{GB} &=& - \fr{\b V_{D-2}}{8 \pi G} 
\ls \fr{8 (D-1)}{8 + (D-2) \et^2} - \fr{8 (D-1) + (D-2) \et^2}{2 \ls 8 + (D-2) \et^2 \rs} \
\fr{r_h^c}{r_0^c}   \rs \  r_0^c , \nn
S_{ct} &=& \fr{\b V_{D-2}}{16 \pi G} 
 \fr{8 (D-2)}{8 + (D-2) \et^2}   \sqrt{1 - \fr{r_h^c}{r_0^c} } \  r_0^c .
\eea
Summing all results, the exact renormalized action finally becomes
\be
S = - \fr{\b V_{D-2}}{16 \pi G} \lb  \fr{16 (D-2)}{8 + (D-2) \et^2} 
\ls 1 - \sqrt{1 - \fr{r_h^c}{r_0^c} }   \rs r_0^c -  
\fr{8 (D-3) + (D-2) \et^2}{8 + (D-2) \et^2} \  r_h^c \rb . 
\ee
For $r_0 \to \inf$, the renormalized on-shell action simply reduces to
\be
S = - \fr{\b V_{D-2}}{16 \pi G} \ \fr{8 - (D-2) \et^2}{8 + (D-2) \et^2}  \  r_h^c .
\ee
Like the AdS/CFT correspondence, the free energy of the dual theory 
can be defined by $F = S/\b$ which gives rise to the same free energy in \eq{thres:freeenergy}.

Now, let us evaluate the boundary energy-momentum tensor. 
From the gravitational action in \eq{act:gravity}, the corresponding boundary stress tensor
defined by $T_{ab} = - \fr{2}{\sqrt{\g}} \fr{\pa {\cal S}}{\pa \g^{ab}}$
reads
\be
{{T_{(bd)}}^a}_b = - \fr{1}{8 \pi G} \int d^{D-2} x \sqrt{\g} \ \g^{ac} \ls \Th_{cb} 
-  \g_{cb} \Th \rs ,
\ee
and the contribution from the counter term is
\be
{{T_{(ct)}}^a}_b= \fr{V_{D-2}}{8 \pi G}  \fr{8 (D-2)}{8 + (D-2) \et^2} \ \sqrt{\g} \ e^{\et \ph /2} \ {\d^a}_b ,
\ee
where the indices, $a$ and $b$, imply the directions of the boundary space and time.
So the renormalized stress tensor is given by the sum of them
\be
{T^a}_b = {{T_{(bd)}}^a}_b +  {{T_{(ct)}}^a}_b .
\ee
The explicit form of the renormalized boundary energy reads
\be
E = {T^{\ta}}_{\ta} = \fr{V_{D-2}}{8 \pi G}  \fr{8 (D-2)}{8 + (D-2) \et^2} 
\lb \sqrt{1 - \fr{r_h^c}{r_0^c} }  - \ls  1 - \fr{r_h^c}{r_0^c}   \rs \rb \ r_0^c ,
\ee
and the pressure becomes
\be
P = - \fr{{T^{i}}_{i}}{V_{D-2}}  = \fr{1}{8 \pi G} \lb \fr{8 (D-2)}{8 + (D-2) \et^2}  
\ls 1 -  \sqrt{1 - \fr{r_h^c}{r_0^c} }  \rs 
- \fr{8 (D-3) + (D-2) \et^2}{2 ( 8 + (D-2) \et^2) }  \fr{r_h^c}{r_0^c} \rb \ r_0^c ,
\ee 
where the pressure is the same as $- F$.  
If we put an UV cutoff into infinity $r_0 = \inf$, the energy and pressure are simply reduced to 
\bea		\la{res:holographicem}
E &=& \fr{V_{D-2}}{8 \pi G}  \fr{4 (D-2)}{8 + (D-2) \et^2}  r_h^c , \nn
P &=& \fr{1}{16 \pi G}  \fr{ 8 - (D-2) \et^2}{8 + (D-2) \et^2}  r_h^c .
\eea
These results show that the holographic renormalization of the Einstein-dilaton theory 
reproduces the exact same EdBB thermodynamics when the boundary is located at infinity.

At zero temperature ($r_h=0$), the free energy and the internal energy of 
the dual theory become zero. 
In \cite{Kulkarni:2012re}, it was shown that for $\et^2 \ge 8/(D-2) $
the EdBB and its dual theory are thermodynamically unstable.
In the parameter range $\et^2 < 8/(D-2)$, the free energy at finite temperature 
is always negative. Therefore, if we do not insert an IR cutoff by hand 
there is no Hawking-Page transition and the EdBB geometry is always preferable. This result is qualitatively the same as one obtained in \cite{Kulkarni:2012in},
where a different renormalization scheme, the so-called subtraction method, was used.  
Now, let us consider the UV cutoff dependence of the free energy. 
If we interpret the position of boundary 
as the energy scale of the dual theory, we can see how the free energy depends on
the energy scale of the dual theory. In Fig 1, we draw the free energy depending on
$r_0$ for $\et = 0$ and $1$, which shows that the free energy monotonically decreases
as the energy scale of the dual theory decreases.

\begin{figure}
\begin{center}
\vspace{0cm}
\hspace{-0.5cm}
\subfigure{ \includegraphics[angle=0,width=0.5\textwidth]{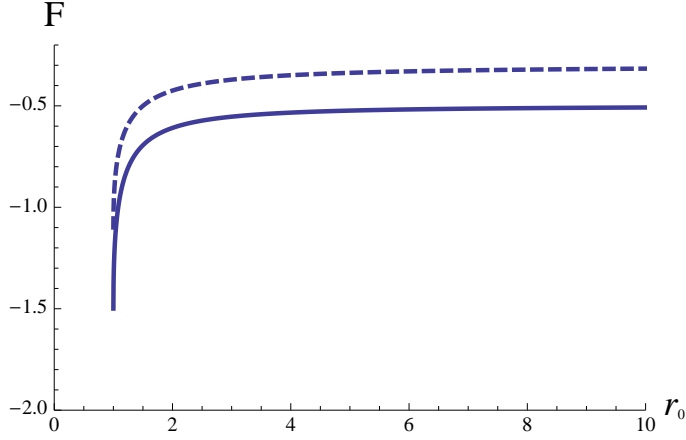}}
\vspace{-0cm} \\
\caption{\small The free energy depending on $r_0$ for $\et=0$ (solid) and $\et=1$ 
(dashed), where we set $8 \pi G = 1$, $r_h = 1$, $D=4$
and $V_{2} = 1$. }
\label{fig1}
\end{center}
\end{figure}

Following the definition of the equation of state parameter, it becomes in the dual relativistic 
non-conformal system
\be
w = \fr{1}{D-2} - \fr{\et^2}{8} ,
\ee
where the second term represents the deviation from the conformal one.
This result also shows that 
the sound velocity of the relativistic non-conformal medium 
\be
c_s  = \sq{w},
\ee 
is always smaller than that of the relativistic conformal one, $\sqrt{\fr{1}{D-2}}$.

\section{Membrane paradigm}

In the previous section, the thermodynamic properties of 
a relativistic non-conformal system has 
been investigated by using the holographic renormalization. 
In this section, we will study the macroscopic properties, especially the
hydrodynamic transport coefficients of the membrane paradigm \cite{Iqbal:2008by,Matsuo:2011fk,Matsuo:2012bv}, and compare them with the results of the Kubo formula \cite{Kulkarni:2012re,Kulkarni:2012in}.

\subsection{Charge diffusion}

In order to describe the charge diffusion process in the relativistic non-conformal medium, we 
need to introduce a Maxwell term describing $U(1)$ gauge field fluctuations on the EdBB. 
Due to the existence of the nontrivial dilaton field, 
the Maxwell term can have a more
general gauge coupling depending on the radius $r$
\be
S= -  \int d^D x \sqrt{-g} \ \frac{1}{4 g_D^2 (r)} \ F_{MN} F^{MN} ,
\ee
with
\be
g_D^2 (r)= {e^{\a \ph(r)}}/{g_0^2} ,
\ee
where $g_0^2$ is a constant and a new parameter $\a$ describes the strength of the gauge 
coupling. 
Usually, the vector fluctuations can be divided into two parts: if the fluctuation moves in the 
$y$-direction with a momentum $k$, one is the longitudinal modes, $A_t$ and $A_y$, 
and the other is the transverse modes $A_x$ (where $x$ means all transverse directions) in the $A_r =0$ gauge. 
Since the charge diffusion process is related to the motion of the longitudinal modes, we 
concentrate only on the longitudinal modes from now on. 
In the hydrodynamic limit
($\o \sim k^2$, $\o << T_H$, and $k << T_H$), the Fourier mode expansions of the longitudinal modes become
\bea
A_t (r,t,y) &=& \int \fr{d \o \ d k}{(2 \pi)^2} \ e^{- i \o t + i k y} A_t (r,\o,k)  , \nn
A_y (r,t,y) &=& \int \fr{d \o \ d k}{(2 \pi)^2} \ e^{- i \o t + i k y} A_y (r,\o,k) .
\eea
These longitudinal modes satisfy two dynamical equations, 
the current conservation and the Bianchi identity. 
In terms of current $j^{\m}$, the governing equations are \cite{Iqbal:2008by}
\bea
0 &=& - \pa_r  j^t - \frac{\sqrt{-g}}{g_D^2 (r)} \ g^{tt} g^{yy} \pa_y F_{yt} , \\
0 &=& - \pa_r  j^y - \frac{\sqrt{-g}}{g_D^2 (r)} \ g^{tt} g^{yy} \pa_t F_{yt} , \\
0 &=&  \pa_t j^t + \pa_y j^y , \la{rel:conservation} \\
0 &=& - \frac{g_{rr} g_{yy} g_D^2 (r) }{\sqrt{-g}} \pa_t  j^y 
- \frac{g_{rr} g_{yy} g_D^2 (r) }{\sqrt{-g}} \pa_z j^t + \pa_r F_{yt}.
\eea
Combining these equations, one can easily derive a flow equation for the longitudinal conductivity 
defined by $\s_L (r,k_{\m}) \equiv j^y / F_{yt}$
\be
\pa_r \s_L = i \o \sqrt{\frac{g_{rr}}{g_{tt}}} \lb \frac{\s_L^2}{\S_A (r)} 
\ls 1 - \frac{k^2}{\o^2} \frac{g^{yy}}{g^{tt}}\rs - \S_A (r) \rb ,
\ee
with
\be
\S_A (r)  = \frac{1}{g_D^2 (r)} \sqrt{\frac{-g}{g_{tt} g_{rr}}} \ g^{yy} .
\ee

In the zero frequency limit, we can see that the longitudinal conductivity reduces to the DC 
conductivity and that it is
independent of the position of the membrane. Furthermore, imposing the 
regularity of the conductivity at the horizon the DC conductivity leads to
\bea		\la{initialcond}
\s_{DC} 
&=& \frac{1}{g_0^2}  \ls \frac{ 4 \pi  [8 + (D-2) \eta^2]}{8 ( D - 1) - (D-2) \eta^2}  
\rs^{\frac{8 ( D - 4) - 2 (D-2) \eta \a}{8 - (D-2) \eta^2}}
T_H^{\frac{8 ( D - 4) - 2 (D-2) \eta \a}{8 - (D-2) \eta^2}}  .
\eea 
For the special values of $\a$, $0$ and $- \eta/2$, the above DC conductivity  
reproduces the coincident results obtained by the Kubo formula \cite{Kulkarni:2012re}.
From the general DC conductivity in \eq{initialcond},
we can see that the DC conductivity of the relativistic non-conformal theory is always positive 
and real because the range of $\et$ 
should be constrained to $\et^2 < \fr{8}{D-2}$. 
Moreover, if $\a > \fr{4 (D-4)}{(D-2) \et}$ the DC conductivity decreases with increasing
temperature. Especially,
taking $\a = \fr{8 (D-1) - (D-2) \et^2}{2 (D-2) \et}$ for $D=5$ provides the resistivity 
proportional to temperature, which is the macroscopic electric property 
of the metal.
If $\a = \fr{4 (D-4)}{(D-2) \et}$, the DC conductivity is independent of temperature.
For $\a < \fr{4 (D-4)}{(D-2) \et}$, it increases with temperature, which is 
a typical feature of the electrolytes or some chemical compounds. 
In general, the macroscopic electric properties 
crucially depend on what the charge carriers are. For example, the electric property of the metal
is mainly governed by the motion of electrons, whereas the motion of ions is important to 
understand the electric properties of the electrolytes or chemical compounds.
As a result, we can say that the parameter $\a$ provides the information for the charge 
carrier in the dual field theory.

In the hydrodynamic limit ($\o, k << T_H$), the flow equation of the longitudinal conductivity reduces to
\be			\la{eq:eqfordiffusion}
\frac{\pa_r \s_L}{\s_L^2} = - i \frac{k^2}{\o} \frac{ g_{tt}  g_{rr}}{\sqrt{-g}}  \ g_D^2 (r)  .
\ee
According to the holographic renormalization, the radial position of the membrane $r_m$ can be identified with the energy scale of the dual field theory. 
Using the fact that the DC conductivity plays a role of the initial data for the conductivity flow
\cite{Iqbal:2008by},
integrating \eq{eq:eqfordiffusion} from $r_h$ to $r_m$ gives rise to  
the following longitudinal conductivity
\be
\s_L (r_m) = \frac{i \o \ \s_{DC}}{ i \o - {\cal D}_e (r_m) k^2} ,
\ee 
where the charge diffusion constant ${\cal D}_e (r_m)$ depending on the energy scale is given by
\bea    \la{res:chargediffcon}
{\cal D}_e (r_m) &=& \s_{DC} \int_{r_h}^{r_m} dr \ \frac{g_{tt}  g_{rr}}{\sqrt{-g}}   \ g_D^2 (r) \nn
&=&  \frac{(  r_h^{-\d}  -  r_m^{-\d} ) }{\d}  \ \ r_h^{\chi},
\eea
with
\bea
\d  &=&  \frac{8 (D-3) - (D - 2)  (\eta + 2 \a ) \eta}{8 + (D-2) \eta^2}, \nn
\chi &=& \frac{8 (D-4) - 2 (D - 2) \a \eta}{8 + (D-2) \eta^2} .
\eea
For a positive $\d$ $\ls {\rm or} \ \a < \frac{8 (D-3) - (D-2) \eta^2}{ 2 (D-2) \eta } \rs$, the charge diffusion constant of the
dual field theory is well-defined in the limit of $r_m \to \infty$. Otherwise, it diverges
at the infinity. For a well-defined hydrodynamic transport coefficients at the 
asymptotic
boundary, we concentrate only on the positive $\d$ from now on. 
Then, the general charge diffusion constant at the asymptotic boundary becomes
\be   \la{res:chargediffrgflow}
{\cal D}_e  (\infty)= \fr{1}{4 \pi T_H} \ \fr{8 (D- 1) - (D-2) \et^2}{8 (D-3) - (D-2)  
(\et + 2 \a  ) \et } .
\ee
In the $4$-dimensional case, the charge diffusion constant at the asymptotic boundary 
is consistent with the result of the Kubo formula \cite{Kulkarni:2012re}.
For more understanding, we take a special value of $\a$, for simplicity $\a=0$. 
Since $\et$ is always smaller than $\fr{8}{D-2}$ due to the thermodynamic stability,
the charge diffusion constant increases with $\et$. This fact implies that 
the charge diffusion constant of the non-conformal medium is larger than that of the conformal
one.
Moreover, since the half life-time of the quasi normal mode in the diffusion process is inversely
proportional to the charge diffusion constant and the square of the momentum, 
the above result also implies that the quasi normal mode decays more rapidly in the non-conformal
medium. 
Lastly, we can easily see from \eq{res:chargediffcon} that the charge diffusion constant 
decreases monotonically as the energy of the dual theory runs from UV to IR (see Fig. 2).

\begin{figure}
\vspace{0cm}
\centerline{\epsfig{file=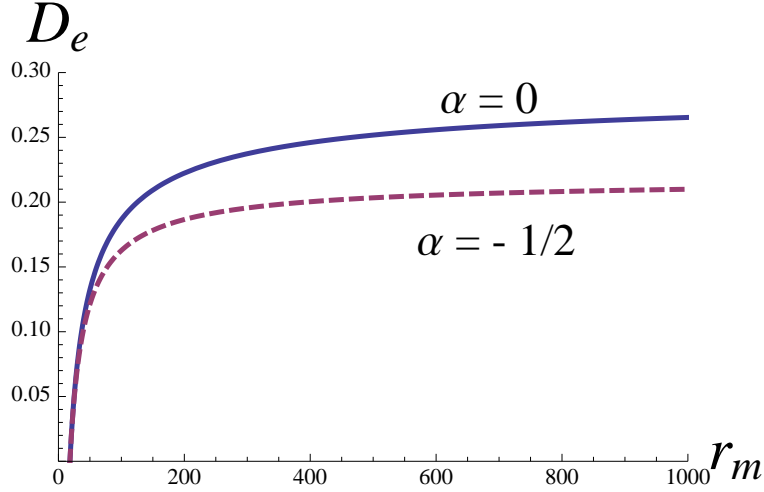,width=10cm}}
\vspace{0cm}
\caption{\small The RG flow of the charge diffusion constant, where we take $\et=1$,
$T_H =1$ and $D=4$. It shows that the charged diffusion constant increases monotonically  
with the energy scale denoted by $r_m$ and it becomes zero at the horizon $r_h = 18.2521$. }
\label{density}
\end{figure}

\subsection{Momentum diffusion}

As shown in \cite{Kovtun:2003wp,Iqbal:2008by}, the relevant equations for gravitational shear modes
$h^x_t$ and $h^x_y$, can
be mapped to an electromagnetic problem. If we set $h^x_t = a_t$ and $h^x_y = a_y$,
the action for the shear modes reduces to
\be
S = \frac{1}{16 \pi} \int d^D x \sqrt{-g}\  g_{xx} F_{\a \b} F^{\a \b} ,
\ee
where $\a$ and $\b$ imply the longitudinal direction $t$ or $y$. 
$F_{\a \b}$ is the field strength of $a_t$ and $a_y$, which in terms of metric fluctuations 
is given by
$F_{\a \b} = \pa_{\a} h^x_{\b} - \pa_{\b} h^x_{\a}$. This action for shear modes  
is exactly the standard Maxwell form with an effective coupling $g_G^2$
\be
\frac{1}{g_G^2} = \frac{1}{16 \pi} \ g_{xx} .
\ee
So we can immediately take over all results of the previous section. 

The counter part of the DC conductivity denoted by $\s_G$ becomes
\be		
\s_{G} = \lp \frac{1}{16 \pi}   \sqrt{\frac{-g}{g_{tt} g_{rr} }}  \right|_{r_h}  .
\ee
Notice that the Bekenstein-Hawking entropy density $s$ can be written as
\be
s \equiv \frac{S_{BH}}{V_{D-2}} =  \lp \frac{1}{4}  \sqrt{\frac{-g}{g_{tt} g_{rr} }}\right|_{r_h} .
\ee
From the celebrated universality in \eq{des:univers}, one can easily see that $\s_G$ is
nothing but the shear viscosity $\et_s$ and that
the result is the expected one in the Kubo formula \cite{Kulkarni:2012in}.

Similarly, we can also easily evaluate the retarded Green function of the shear modes
\be
G_R^{xy,xy} = \frac{\eta_s \ \o^2}{i \o - D_s (r_m) \ k^2},
\ee
where the momentum diffusion constant ${\cal D}_s (r_m) $, by taking the analogy to the charge diffusion
constant, is 
\bea
{\cal D}_s (r_m) &=&   4 s \int_{r_h}^{r_m} dr \frac{g_{tt} \ g_{rr}}{\sqrt{-g} \ g_{xx}} \nn
&=& \frac{ 8+(D-2) \eta^2 }{8(D-1) - (D-2) \eta^2}  \ls \frac{1}{r_h^{\g}} - \frac{1}{r_m^{\g}} \rs .
\eea
with
\be
\g = \frac{8 - (D-2) \eta^2}{8 + (D-2) \eta^2} .
\ee
Below the crossover value $\eta^2 < 8/(D-2)$ \cite{Gursoy:2007cb,Gursoy:2007er,Gouteraux:2011ce} 
where the black brane is thermodynamically stable, $\g$ is always positive. If we put
the membrane at the infinity and rewrite the momentum diffusion constant in terms of
temperature, the momentum diffusion constant has the form expected by the 
EdBB thermodynamics
\be
{\cal D}_s (\infty) = \frac{1}{4 \pi T_H} ,
\ee 
which shows the self-consistency of the gauge/gravity duality in the relativistic non-conformal medium.
In addition, this result implies that the corresponding quasi normal mode decays rapidly at high temperature, whereas the momentum diffusion constant
does not depend on the non-conformality unlike the charge diffusion constant.
Finally, we plot the momentum diffusion constant depending on the energy scale in Fig. 3.
It shows,  similar to the charge diffusion constant, the monotonically decreasing behavior 
with decreasing energy.

\begin{figure}
\vspace{0cm}
\centerline{\epsfig{file=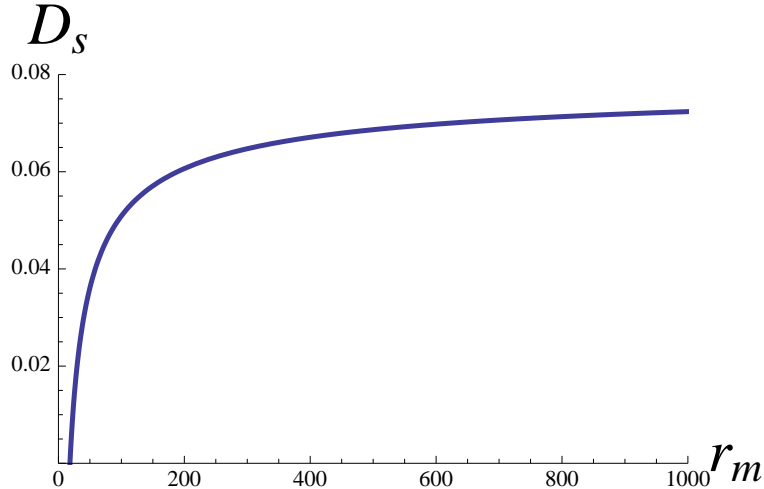,width=10cm}}
\vspace{0cm}
\caption{\small The RG flow of the momentum diffusion constant, where we take $\et=1$,
$T_H =1$ and $D=4$. Similarly to the charged diffusion constant, the momentum diffusion constant also increases monotonically with the energy scale $r_m$ and it becomes zero at the
horizon $r_h = 18.2521$. }
\label{density}
\end{figure}

\section{Binding energies of particles and monopoles}

In the holographic QCD, the binding energy of quark and antiquark is described by a 
temporal Wilson loop, which in the string theory corresponds to the trajectory of the open string ends. In this section, we will investigate such a binding energy in the relativistic non-conformal 
medium. Before starting the calculation, it is worth to note that there is no difference between 
a fundamental string (or F1-string) and D1-brane (or D1-string) in the AdS space if we ignore the gravitational backreaction of the them. 
However, that is not true in the EdBB geometry due to the nontrivial dilaton profile.
The metric usually felt by an open string is not in the Einstein frame but rather in the string frame represented as 
\cite{Gubser:1999pk,Bak:2004yf,Kim:2008ax}
\be
ds^2_{string} = e^{\fr{\ph}{2}} \ ds^2_{Einstein} ,
\ee
where the Einstein metric is given by \eq{eq:asymp}. Due to the
nontrivial dilaton field, the string actions for them usually have the different $\ph$-dependence. 
Since the end of an F1- and D1-string corresponds to a particle and monopole respectively,
we can easily expect that the binding energies of particles and monopole are different
in the relativistic non-conformal theory.
So, the goal of this section is to investigate the binding energies of particles and monopoles
as well as to study the effect of the non-conformality on them.

In order to investigate the binding energy of two fundamental particles, we
take into account the Nambu-Goto action of a fundamental string 
\be
S_{F1} = \fr{1}{2 \pi \a'} \int d^2 \s \sqrt{- \det \ls G_{\m \n} \fr{\pa x^{\m}}{\pa \s^{\a}} 
\fr{\pa x^{\n}}{\pa \s^{\b}} \rs}  ,
\ee
where $G_{\m \n}$ is the space-time metric in the string frame. 
In the static gauge with the metric in \eq{eq:asymp}
\be		\la{par:f1}
\ta = t, \quad \s = x^1 = x, \quad x^2 = \cdots = x^{D-2} = 0   \ {\rm and}  \quad r = r(x) ,
\ee
assuming that the end points of string are located at 
$\lc r, x \rc = \lc \infty, \pm l/2 \rc$,
then the string action simply reduces to
\be
S_{F1} = \fr{\b}{2 \pi \a'} \int_{- l /2}^{l/2} d x \ e^{\ph /2} \sqrt{ \dot{r}^2 
+ r^{4 a_1} f(r) } ,
\ee
where $\b$ is the time interval and dot means a derivative with respect to $x$.
Using the analogy to mechanics, the conserved Hamiltonian after 
regarding $x$ as time is given by
\be
H = - \fr{e^{\ph /2} }{2 \pi \a'} \fr{r^{4 a_1} f(r)}{\sqrt{ \dot{r}^2 
+ r^{4 a_1} f(r) }} .
\ee
If $r$ has a turning point or minimum value $r_*$ satisfying $\dot{r}_*= 0$,
the existence of such a turning point implies that particle and antiparticle are connected by a string which corresponds to the bound state of particles. The absence of no turning point
says that an open string connecting two particles divides into two straight strings describing free particles. 
First, we concentrate on the string configuration with a turning point at which the 
above conserved Hamiltonian is still satisfied
\be
H = - \fr{e^{\ph_* /2} }{2 \pi \a'} \ r_*^{2 a_1} \sqrt{f(r_*)} ,
\ee 
with $\ph_* = \ph(r_*)$.

After introducing a rescaled coordinate $\td{r}= r/r_*$, comparing above two Hamiltonians gives rise to information for the interdistance between particles in terms of the turning point 
\be
l = \fr{2 \sqrt{f(1)} }{r_*^{2 a_1-1}}  \int_1^{\inf} d \td{r} \fr{1}{\td{r}^{2 a_1} \sqrt{f(\td{r})}
\sqrt{e^{\td{\ph}} \ \td{r}^{4 a_1} f(\td{r}) - f (1) }} ,
\ee
where $f(\td{r}) = 1 - \td{r}_h^c/\td{r}^c$, $f(1) = 1 - \td{r}_h^c$ and $\td{\ph} = \ph(\td{r})$.  In addition, an unrenormalized energy of a pair of particles
becomes
\be
E \equiv \fr{S_{F1}}{\b} = \fr{r_*^{1-\fr{k_0}{2}}}{\pi \a'} \int_1^{\inf} d \td{r} \fr{e^{\td{\ph}} \ \td{r}^{2 a_1} \sqrt{f(\td{r})} }{\sqrt{e^{\td{\ph}} \ \td{r}^{4 a_1} f(\td{r}) - f (1)  }} .
\ee
In the asymptotic region ($\td{r} \to \inf$), the unrenormalized energy has the following 
approximate form
\be
E  \approx   \fr{r_*^{1-\fr{k_0}{2}}}{\pi \a'} \fr{\td{r}^{1-k_0/2}}{1- k_0/2}.
\ee
Since $1- k_0/2 > 0$ for $\et^2 < \fr{8}{D-2}$, 
the unrenormalized energy diverges when $\td{r} \to \infty$.
In order to define the binding energy well, we need to renormalize it by adding an appropriate counter term.
In the holographic QCD, this kind of divergence appears due to the infinite masses of two quarks
described by straight strings. Therefore, we can remove the above divergence by 
subtracting the infinite particle masses. To do so, we parameterize two straight strings as
\be			\la{par:fcounter}
\ta = t , \quad \s = r , \ {\rm and } \ x^1 = \pm \fr{l}{2} .
\ee
Then, the energy of two straight strings is given by
\be
E_{ct}  = \fr{r_*^{1-\fr{k_0}{2}}}{\pi \a'} \int_{\td{r}_h}^{\inf} d \td{r} 
\ e^{\td{\ph}/2} ,
\ee
which corresponds to the rest mass of two particles and can exactly cancel the
divergence of the unrenormalized energy. 
The renormalized energy of a pair of particles becomes
\bea
V &\equiv& E - E_{ct}  \nn
&=& \fr{r_*^{1-\fr{k_0}{2}}}{\pi \a'} \ls \int_1^{\inf} d \td{r} \fr{e^{\td{\ph}} r^{2 a_1} 
\sqrt{f(\td{r})}}{\sqrt{e^{\td{\ph}} \ r^{4 a_1} f(\td{r}) - f (1) }} 
 - \int_{\td{r}_h}^{\inf} d \td{r} 
\ e^{\td{\ph}/2} \rs ,
\eea
which corresponds to the well-defined binding energy between particle and antiparticle.

Now, let us consider the low temperature case ($\td{r}_h = r_h / r_* \ll 1$), in which 
the interdistance and the binding energy of a pair of particle have the following expansion forms
\bea
l &=& \fr{1}{r_*^{2 a_1 - 1}} \ls A_0 + A_1  \fr{r_h^c}{r_*^c} + \cdots \rs
, \la{rel:distance} \\
V &=& B_0 \ r_*^{1 - \fr{k_0}{2}} + B_1 \ r_*^{1 - \fr{k_0}{2}-c} \ r_h^c + B_2 \ r_h^{1 - \fr{k_0}{2}}
+ \cdots  ,
\la{rel:binding}\eea
where ellipsis means higher order corrections and
\bea
A_0 &=& \fr{2 \sqrt{\pi} }{2 a_1 -1 } 
\fr{\G \ls \fr{1}{2} + \fr{2 a_1 -1}{4 a_1 - k_0} \rs}{\Gamma 
\ls \fr{2 a_1 -1}{4 a_1 - k_0} \rs} , \nn
A_1 &=& \fr{2 \sqrt{\pi} }{4 a_1 - k_0}   
\lb  \fr{\Gamma \ls \half + \fr{2 a_1 -1}{4 a_1 - k_0} \rs}{\Gamma \ls \fr{2 a_1-1}{4 a_1 - k_0}\rs} + \fr{2  - 2c  - k_0}{2 (4 a_1 - k_0)} \ 
\fr{\Gamma \ls \half + \fr{2 a_1 -1 + c}{4 a_1 - k_0} \rs }{\Gamma \ls 1 + \fr{2 a_1 -1 + c}{4 a_1 - k_0} \rs} \rb
 , \nn
B_0 &=& - \fr{2}{(2 - k_0) \sqrt{\pi} \a' } \fr{\Gamma \ls \half + \fr{2 a_1 -1}{ 4 a_1 -  k_0} \rs}
{\Gamma \ls \fr{2 a_1 -1}{4 a_1 - k_0} \rs}, \nn
B_1 &=&  \fr{1}{ (4 a_1 - k_0) \sqrt{\pi} \a' }  \lb 
\fr{\Gamma \ls \half + \fr{2 a_1 -1}{4 a_1 - k_0} \rs}
{\Gamma \ls \fr{2 a_1 -1}{4 a_1 - k_0} \rs}
- \fr{\Gamma \ls \half + \fr{ 2 a_1 -1 + c}{4 a_1 - k_0} \rs}{\Gamma \ls \fr{
2 a_1 -1 + c}{ 4 a_1 - k_0 } \rs} \rb , \nn
B_2 &=& \fr{1}{\pi \a' \ls 1 - \fr{k_0}{2} \rs }   . 
\eea

In the zero temperature limit ($r_h \to 0$), the binding energy shows 
a Coulomb-like potential with a power depending on the non-conformality
\be
V = \fr{A_0^{\g} \ B_0 }{ l^{ \g} },
\ee
with 
\be
\g = \fr{8 - (D-2) \et + (D-2) \et^2}{8 - (D-2) \et^2} .
\ee
For the conformal theory ($\et=0$) dual to the AdS space, the binding energy 
is given by the Coulomb potential proportional to $l^{-1}$, as expected by the conformality.
For $0 < \et < 1/2$, $\g$ is positive and smaller than $1$. This implies that,
when comparing with the conformal case, the 
magnitude of the binding energy in the non-conformal medium slowly decreases as the interdistance of two particles increases. Interestingly, the Coulomb potential inversely
proportional to the interdistance
again appears at $\et = 1/2$. Finally, for  $1/2< \et < \fr{2 \sqrt{2}}{D-2}$ the binding energy is steeper than one of the conformal case.

In order to investigate the thermal correction at low temperature, we need to rewrite $r_*$ in terms of $l$ and $r_h$. To do so, let us set
\be			\la{ans:rstar}
r_* = \ls \fr{A_0}{l} \rs^{\fr{1}{2 a_1 -1}}  \ls 1 + \d\rs ,
\ee 
where $\d$ corresponds to the first thermal correction and is a function of $l$ and $r_h$.
Since \eq{ans:rstar} should satisfy \eq{rel:distance} at least at the first order of correction,
$\d$ must be
\be
\d = \fr{1}{2 a_1 -1} \ A_0^{-\fr{2a_1 -1 +c}{2 a_1-1}} \ A_1  \ l^{\fr{c}{2 a_1 -1}} \ r_h^c.
\ee
Inserting this result together with \eq{ans:rstar} into \eq{rel:binding} gives rise to
\be
V = \fr{A_0^{\g} B_0}{l^{\g}} \lb 1 + K \ l^{\fr{c}{2 a_1 - 1}} r_h^c \rb 
\ee
where $K$ is given by
\be
K = \fr{2- k_0}{2(2 a_1 -1)} \ A_0^{- \fr{2a_1 -1 +c}{2 a_1-1}}  \
A_1 + \fr{B_1}{B_0} \ A_0^{- \fr{c}{2 a_1 -1}} ,
\ee
and $r_h^c$ is related to temperature
\be
r_h^c = \ls \fr{4 \pi \lc 8 + (D-2) \et^2 \rc}{8 (D-1)+(D-2) \et^2} \rs^{\fr{8 (D-1)- (D-2) \et^2}{8 - (D-2) \et^2}} 
T^{\fr{8 (D-1)- (D-2) \et^2}{8 - (D-2) \et^2}} .
\ee
As a result, the first thermal correction to the binding energy is proportional to
\be		\la{res:thecorrpar} 
V_T \sim  l^{\fr{(D-2) (8 + \et - 2 \et^2)}{8 - (D-2) \et^2}}  T^{\fr{8 (D-1)- (D-2) \et^2}{8 - (D-2) \et^2}}  .
\ee
This result shows that the thermal correction to the binding energy nontrivially depends
on the interdistance and temperature with the power determined by the non-conformality
and dimension. 
Especially, for the conformal case $\et=0$ the first thermal correction is proportional to
\be
V_T \sim l^{D-2} \ T^{D-1}.
\ee 
At finite temperature, the binding energy of particles is calculated numerically
in Fig. 4. This result shows that the magnitude of the binding energy increases with increasing non-conformality $\et$.

\begin{figure}
\begin{center}
\vspace{0cm}
\hspace{-0.5cm}
\subfigure{ \includegraphics[angle=0,width=0.5\textwidth]{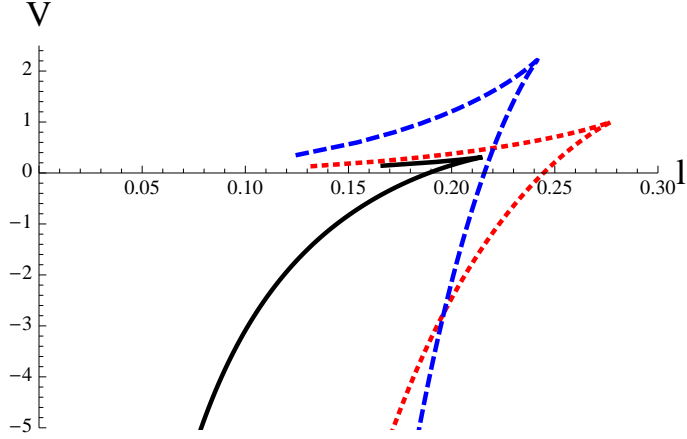}}
\vspace{-0cm} \\
\caption{\small At given temperature $T=1$ the binding energy of particles for $\et=0$ (solid), which is the same as the one of monopoles. The dashed or dotted one represent
the binding energies of particles or monopoles for $\et=1$ respectively, where we set $\pi \a' = 1$ and $D=4$. }
\label{fig2}
\end{center}
\end{figure}

It is also possible to think of the binding energy for a pair of monopole and anti-monopole
by considering a $D1$-string instead of an $F1$-string. In the dual theory, the end of
$D1$-string correspond to a monopole or anti-monopole whereas the end of a fundamental
string describes a fundamental particle. In the AdS background, since there is no
nontrivial dilaton profile, the binding energy of a pair of monopole and anti-monopole
is the same as one for particles. However, there exists a nontrivial 
dilaton field in the dual geometry of the relativistic non-conformal theory, so the binding
energy is different from the $F1$-string result.  
In order to investigate the binding energy of a pair of monopole and anti-monopole, we
should consider a Nambu-Goto action of a $D1$-string in the string frame, which contains an extra dilaton field contribution
\bea
S_{D1} &=& \fr{1}{2 \pi \a'} \int d^2 \s \ e^{- \ph} \sqrt{- \det \ls G_{\m \n} \fr{\pa x^{\m}}{\pa \s^{\a}} 
\fr{\pa x^{\n}}{\pa \s^{\b}} \rs} \nn
 &=& \fr{\b}{2 \pi \a'} \int_{- l /2}^{l/2} d x \ e^{- \ph /2} \sqrt{ \dot{r}^2 
+ r^{4 a_1} f(r) } ,
\eea
where the same parameterization \eq{par:f1} is used.
In terms of the dimensionless coordinate $\td{r}$, the interdistance of two monopoles 
can be rewritten as
\be
l = \fr{2 \sqrt{f(1)}}{r_*^{2 a_1-1}}   \int_1^{\inf} d \td{r} \fr{1}{\td{r}^{2 a_1} \sqrt{f(\td{r})}
\sqrt{e^{- \td{\ph}} \ \td{r}^{4 a_1} f(\td{r}) - f (1) }} ,
\ee
where $f(\td{r}) = 1 - \td{r}_h^c/\td{r}^c$ and $\td{\ph} = \ph(\td{r})$, and the 
unrenormalized energy is 
\be
E = \fr{r_*^{1+\fr{k_0}{2}}}{\pi \a'} \int_1^{\inf} d \td{r} \fr{e^{-\td{\ph}} \ \td{r}^{2 a_1} \sqrt{f(\td{r})} }{\sqrt{e^{-\td{\ph}} \ \td{r}^{4 a_1} f(\td{r}) - f (1)  }} .
\ee
Similar to the $F1$-string the divergence of the above unrenormalized energy 
can be renormalized by the following counter term, which corresponds to infinitely massive two
monopoles described by two straight $D1$-strings,
\be
E_{ct}  = \fr{r_*^{1+\fr{k_0}{2}}}{\pi \a'} \int_{\td{r}_h}^{\inf} d \td{r} 
\ e^{- \td{\ph}/2} ,
\ee
where the same parameterization in \eq{par:fcounter} is used.
Then, the resulting renormalized binding energy of a pair of monopole and anti-monopole 
reduces to
\be
V 
= \fr{r_*^{1+\fr{k_0}{2}}}{\pi \a'} \ls \int_1^{\inf} d \td{r} \fr{e^{- \td{\ph}} r^{2 a_1} 
\sqrt{f(\td{r})}}{\sqrt{e^{-\td{\ph}} \ r^{4 a_1} f(\td{r}) - f (1) }} 
 - \int_{\td{r}_h}^{\inf} d \td{r}  \ e^{-\td{\ph}/2} \rs .
\ee

At low temperature ($\td{r}_h = r_h / r_* \ll 1$),
the interdistance and energy of a pair of monopoles have the following expansion forms
\bea
l &=& \fr{1}{r_*^{2 a_1 - 1}} \ls C_0 + C_1  \fr{r_h^c}{r_*^c} + \cdots \rs
, \la{rel:mdistance} \\
V &=& D_0  \ r_*^{1 + \fr{k_0}{2}} + D_1 \ r_*^{1 + \fr{k_0}{2}-c} \ r_h^c + 
D_2 \ r_h^{1 + \fr{k_0}{2}} + \cdots ,
\la{rel:mbinding}\eea
where
\bea
C_0 &=& \fr{2 \sqrt{\pi} }{2 a_1 -1 } 
\fr{\G \ls \fr{1}{2} + \fr{2 a_1 -1}{4 a_1 + k_0} \rs}{\Gamma 
\ls \fr{2 a_1 -1}{4 a_1 + k_0} \rs} , \nn
C_1 &=& \fr{2 \sqrt{\pi} }{4 a_1 + k_0}   
\lb  \fr{\Gamma \ls \half + \fr{2 a_1 -1}{4 a_1 + k_0} \rs}{\Gamma \ls \fr{2 a_1-1}{4 a_1 + k_0}\rs} + \fr{2  - 2c  + k_0}{2 (4 a_1 + k_0)} \ 
\fr{\Gamma \ls \half + \fr{2 a_1 -1 + c}{4 a_1 + k_0} \rs }{\Gamma \ls 1 + \fr{2 a_1 -1 + c}{4 a_1 + k_0} \rs} \rb
 , \nn
D_0 &=& - \fr{2}{(2 + k_0) \sqrt{\pi} \a' } \fr{\Gamma \ls \half + \fr{2 a_1 -1}{ 4 a_1 +  k_0} \rs}
{\Gamma \ls \fr{2 a_1 -1}{4 a_1 + k_0} \rs}, \nn
D_1 &=&  \fr{1}{ (4 a_1 + k_0) \sqrt{\pi} \a' }  \lb 
\fr{\Gamma \ls \half + \fr{2 a_1 -1}{4 a_1 + k_0} \rs}
{\Gamma \ls \fr{2 a_1 -1}{4 a_1 + k_0} \rs}
- \fr{\Gamma \ls \half + \fr{ 2 a_1 -1 + c}{4 a_1 + k_0} \rs}{\Gamma \ls \fr{
2 a_1 -1 + c}{ 4 a_1 + k_0 } \rs} \rb , \nn
D_2 &=& \fr{1}{\pi \a' \ls 1 + \fr{k_0}{2} \rs }.
\eea
Similar to \eq{ans:rstar}, $r_*$ can be rewritten in terms of $l$ and $r_h$ as
\be
r_* =  \ls \fr{C_0}{l} \rs^{\fr{1}{2 a_1 -1}}  \ls 1 +  
\fr{1}{2 a_1 -1} \ C_0^{-\fr{2a_1 -1 +c}{2 a_1-1}} \ C_1  \ l^{\fr{c}{2 a_1 -1}} \ r_h^c\rs .
\ee
Inserting this result into the binding energy, we finally obtain
\be
V = \fr{A_0^{\chi} B_0}{l^{\chi}} \lb 1 + L \ l^{\fr{c}{2 a_1 - 1}} r_h^c \rb 
\ee
where $\chi$ and $L$ are given by
\bea
\chi &=&  \fr{8 + (D-2) \et + (D-2) \et^2}{8 - (D-2) \et^2} , \nn
L &=& \fr{2 + k_0}{2(2 a_1 -1)} \ C_0^{- \fr{2a_1 -1 +c}{2 a_1-1}}  \
C_1 + \fr{D_1}{D_0} \ C_0^{- \fr{c}{2 a_1 -1}} .
\eea
As a result, the first thermal correction is proportional to
\be
V_T \sim \ls l \  T \rs^{\fr{8 (D-1)- (D-2) \et^2}{8 - (D-2) \et^2}} .
\ee
Unlike the binding energy of particles, the first thermal correction of monopoles 
has the same dependence on the interdistance and temperature.
Similar to the particle case, the magnitude of the binding energy of monopoles at finite temperature 
increases as the non-conformality increases (see Fig. 4).

\section{Drag force}

In this section, we will investigate the drag force of an external particle and a monopole in the 
non-conformal medium. As mentioned before, the action describing the motion of an $F1$- 
or $D1$-string is defined in the string frame \cite{Gubser:2006bz}. 
In the Einstein-dilaton theory, due to the nontrivial dilaton
profile the Nambu-Goto action has an additional contribution from the dilaton field unlike the 
$AdS$ case. In the static gauge 
\be
\ta=t \ , \ \s= r \ {\rm and} \ x_1 = v t + x(r) , 
\ee
the Nambu-Goto action reduces to
\be
S = \frac{1}{2 \pi \a '} \int d^2 \s \ e^{\zeta \ph/2} \sqrt{1 - \fr{v^2}{f} 
+ r^{4 a_1} f x'^2 } .
\ee
where the prime means the derivative with respect to $r$ and 
$\z$ is $+1$ or $-1$ for an $F1$- or $D1$-string respectively\footnote{In
\cite{Kiritsis:2012ta,Alishahiha:2012cm}, the drag force in the general hyperscaling violation background was investigated with $\z=0$.}. The conserved quantity, when regarding $r$ as a time, is represented as
\be
\Pi_x = e^{\z \ph /2} \fr{r^{4 a_1} f x'}{\sqrt{1 - v^2 / f + r^{4 a_1} f x'^2}} ,
\ee
where we set $2 \pi \a ' = 1$.
Rewriting $x'$ as a function of $\Pi_x$ gives rise to
\be
x' = \fr{\Pi_x}{ r^{4 a_1} f} \sqrt{\fr{f-v^2}{e^{\z \ph} f - \Pi_x^2 / r^{4 a_1}}} .
\ee
For a well-defined $x'$, the inside of the square root should be always positive. However,
there exists a point $r_s$ at which $f(r_s) = v^2$ is saturated, so the denominator should
also change its sign at that point. In terms of $r_h$, $r_s$ becomes
\be		\la{res:crtpt}
r_s = \fr{r_h}{(1-v^2)^{1/c}} ,
\ee
and the conserved momentum is expressed by
\be		\la{res:consquan}
\Pi_x = \fr{v}{(1-v^2)^{\d/c}} \ r_h^{\d/2} ,
\ee
where $\d = 4 a_1 - \z k_0$.

The momentum flow along the string is represented by \cite{Gubser:2006bz}
\be 			\la{eq:integraldrag}
\D P_1 =  \int d t  \ \sqrt{-g} \ {P^r}_{x^1} = \fr{d p_1}{dt} \ \D t .
\ee
where the worldsheet current ${P^{\a}}_{\m} $ carried by a string with a nontrivial
dilaton field is given by
\be
{P^{\a}}_{\m}   =  -  e^{\z \ph /2} \ G_{\m  \n}  \ \pa^{\a} x^{\n} .
\ee
Then, the drag force $\fr{d p_1}{d t}$ reads
\be
\fr{d p_1}{dt} = - \sqrt{-g}  \ e^{\z \ph /2} \ G_{x^1  \n} \ g^{r \a}  \ \pa_{\a} x^{\n} ,
\ee
where $G_{x^1  \n}$ is the metric of the target space-time and $g^{r \a}$ is the inverse
of the induced worldsheet metric. 
Using \eq{res:crtpt} and \eq{res:consquan}, the drag force becomes
\be		\la{res:decreasingdrag}
\fr{d p_1}{dt} = - \fr{r_h^{\d /2} \ v }{ (1 - v^2)^{\d / 2 c}  } ,
\ee
where the integral \eq{eq:integraldrag} is evaluated at the asymptotic boundary.
This result guarantees that the momentum decreases as time evolves.
In particular, for the $AdS_5$ space where $\et=\z=0$ it reproduces the result of 
\cite{Gubser:2006bz}.
Rewriting the above result in terms of physical quantities leads to
\be
\fr{d p_1}{dt} = -  F(T_H) \ \fr{p_1  (m^2 + p_1^2)^{\fr{\d - c}{2c} } }{m^{\fr{\d}{c}}} ,
\ee
with 
\be
F(T_H) = 
\ls \fr{4 \pi (8 + (D-2) \et^2) }{8  (D-1) + (D-2) \et^2}  
\rs^{\fr{16 - (D-2)  \z  \et}{8 - (D-2) \et^2}} \
T_H^{\fr{16 - (D-2) \z  \et}{8 - (D-2) \et^2} }
\ee
where $m$ and $p_1$ is the mass and the momentum of a particle or monopole depending on the
value of $\z$. 

In the non-relativisitic limit ($m \gg p_1$), the momentum decreases exponentially 
\be		\la{res:nonrel}
p_1 (t) = p_1 (t_0) \ e^{- \fr{F(T_H)}{m} t}  ,
\ee 
where $p_1 (t_0)$ is the momentum at $t = t_0$.
The above result also shows that the drag force of a monopole ($\z = -1$) increases more rapidly than the one of a particle ($\z = +1$) as temperature increases.
In the relativistic case ($m \ll p_1$), the momentum decreases in power-law, 
\be
p_1 (t) = \lb  p_0^{\fr{c -\d}{c}} - \fr{c-\d}{c} \fr{F(T_H)}{m^{\d /c}}  t 
\rb^{\fr{c}{c-\d}} ,
\ee
where $p_0$ expresses the momentum at $t_1 =0$. For more understanding, if
rewriting $\fr{c}{c-\d}$ in terms of intrinsic parameters
\be
\fr{c}{c- \d} = \fr{8 (D-1) - (D-2) \et^2}{8 (D-5) +  2 (D-2) \z \et - (D-2) \et^2} .
\ee 
it reduces to
\bea
\fr{c}{c- \d} &=& \fr{24 - 2 \et^2}{- 8 +  4 \z \et - 2 \et^2}  \quad {\rm for} \ D=4, \nn
\fr{c}{c- \d} &=& \fr{32 - 3 \et^2}{- 6 \z \et + 3 \et^2}  \quad {\rm for} \ D=5 .
\eea
Assuming that $\et$ is positive, the thermodynamically stable parameter range of $\et$ 
for $D=4$ is given by $0 \le \et < 2$. In this parameter range, $\fr{c}{c-\d}$ is
always negative regardless of $\z$. Therefore, the momenta of a particle and a monopole
decreases as the inverse power of time 
\be		\la{res:invpowlaw}
p_1 (t) = \fr{1}{\lb  \fr{1}{p_0^{\g}} + \fr{F(T_H)}{\g \ m^{\d /c}}  \ t 
\rb^{\g} } ,
\ee
where $\g = \fr{c}{\d -c} > 0$.
In the thermodynamically stable parameter range for $D=5$, $\fr{c}{c-\d}$ is always
negative for a particle and positive for a monopole. So the momentum of a monopole
gives rise to 
\be
p_1 (t) = \lb  p_0^{\fr{c -\d}{c}} - \fr{c-\d}{c} \fr{F(T_H)}{m^{\d /c}} \ t 
\rb^{\fr{c}{c-\d}} ,
\ee
while a particle shows the inverse power law behavior in
\eq{res:invpowlaw}.
In all cases, the momentum decreases more rapidly at high temperature.

\section{Holographic entanglement entropy}

Recently, there was an interesting conjecture that in a small subsystem 
the entanglement temperature has a universal feature 
proportional to the inverse of the size $l$ \cite{Bhattacharya:2012mi}
\be
T_{en} \sim \fr{1}{l} .
\ee
In this section, we will check such a universal feature 
in the relativistic non-conformal theory.
Before doing that, we first check whether the EdBB can provide a consistent dual geometry
or not following \cite{Dong:2012se}. To do so, it is more convenient to rewrite the metric as the hyperscaling violation form. 
After introducing  
\be
r \to (2 a_1 - 1)^{\fr{1}{1-a_1}} u^{- \fr{1}{2 a_1 -1}} \quad {\rm and} \quad 
\lc t, \ x^i \rc \to  (2 a_1 - 1)^{- \fr{a_1}{1 - a_1}}   \lc t, \ x^i \rc,
\ee
the metric becomes
\be			\la{res:metricinu}
ds^2 = u^{- \fr{2 (D-2 - \th)}{D-2}} \ls - f(u) dt^2 + \fr{du^2}{f(u)} + \d_{ij} dx^i dx^j \rs ,
\ee
where the black brane factor $f(u)$ is
\be
f(u) = 1 - \ls \fr{u}{u_h} \rs^{\fr{c}{2 a_1 - 1}} .
\ee
Here, the hyperscaling violation exponent $\th$ is given by
\be 		\la{res:rangeth}
\th 
= - \fr{(D - 2)^2 \et^2}{8 - (D - 2) \et^2}.
\ee
Now, let us concentrate on the symmetry of the asymptotic geometry. 
Since the boundary is located at $u=0$ in the new coordinate,
the asymptotic metric reduces to 
\be    \la{met:hypersc}
ds^2 =  u^{- \fr{2 (D-2 - \th)}{D-2}} \ls   -  dt^{2} + du^{2} +\d_{ij} dx^i dx^j   \rs .
\ee
Under the following scaling transformation
\be
t \to \l t \ , \quad u \to \l u \ , \ {\rm and} \quad   x^i \to\l x^i ,
\ee
the metric transforms as $d s^2 \to \l^{2 \th/(D-2)} d s^2$,
in which the nonzero value of $\th$
indicates the breaking of the conformal symmetry of the dual field theory. 
Nevertheless, the rotational and translational symmetries of the boundary space 
represent that the dual theory is still relativistic. 
As a result, the dual theory of the EdBB
geometry maps to a relativistic non-conformal field theory.    
In this case, the null energy condition reads
\be
\th \lb  (D-2) - \th \rb  \le 0 .
\ee
For a consistent gravity dual, $\th$ should satisfy this null energy condition \cite{Dong:2012se}.
Since $\th$ in \eq{res:rangeth} satisfies the null energy condition for all ranges of $\et$
\bea
\th &\le& 0  \qquad \qquad {\rm for} \ \ \et^2  < 8/(D-2) , \nn
\th &\ge& D-2 \qquad {\rm for}  \ \ \et^2 > 8/(D-2)  ,
\eea
the EdBB geometry \eq{met:hypersc} is a consistent gravity dual of a relativistic non-conformal field theory.  
Although the entire ranges of $\et$ provide a consistent gravity dual, 
only the range, $\et^2  < 8/(D-2)$, is thermodynamically stable.

Now, let us study the holographic entanglement entropy of such a relativistic non-conformal theory by using a $D-2$-dimensional strip \cite{Ryu:2006bv,Ryu:2006ef}. 
A strip in the EdBB background can be parameterized by
\be		\la{ans:strip}
-\fr{l}{2} \le x^1 \le \fr{l}{2} \quad {\rm and} \quad 0 \le x^i \le L \ \ (i= 2, 3, \cdots, D-2) ,
\ee 
where $L$ corresponds to the interval of $x^i$ and we assume that $l \ll L$. 
Since the strip is extended in the radial direction $u$, its  
profile can be represented as a function of $x^1$, $u = u (x^1)$, with the following 
boundary conditions
\be
\e = u \ls - \fr{l}{2} \rs = u \ls \fr{l}{2} \rs,
\ee 
where $\e$ is an appropriate UV cut off of the radial coordinate.
The area of the strip then becomes
\bea
A = 2 L^{D-3} \int_{0}^{l/2} d x^1 \ u^{- (D - 2 - \th)} \sqrt{\fr{u'^2}{f(u)} + 1} ,
\eea
where the prime implies a derivative with respect to $x^1$. 
If regarding $x^1$ as a time,
the conserved energy density is given by
\be
H = - \fr{2 }{u^{D-2 - \th} \sqrt{\fr{u'^2}{f(u)} + 1}} .
\ee
Now, let us assume that the strip configuration has a maximum value, $u_{max}$, which 
corresponds to the turning point or tip of the U-shape configuration in the $x^1$-$u$ plane. 
At the turning point, 
since $u'$ vanishes, the conserved energy density reduces to
\be
H = - \fr{2 }{u_{max}^{D-2 - \th} } .
\ee 
Comparing two conserved energy densities we can represent the distance $l$ and 
the area $A$ of strip in terms of $u_{max}$
\bea
l &=& 2 u_{max} \int_{\e}^1 \fr{\td{u}^{D-2 - \th} \ d \td{u}}{\sqrt{f(\td{u})} 
\sqrt{1 - \td{u}^{2 (D-2 - \th)}}} , \nn
A &=& - \fr{2 L^{D-3}}{u_{max}^{D-3-\th}} \int_{\e}^1 \fr{\ d \td{u}}{
\td{u}^{D-2 - \th}  \sqrt{f(\td{u})} \sqrt{1 - \td{u}^{2 (D-2 - \th)}}} , \la{for:entent}
\eea
where the new coordinate $\td{u}$ is defined as $\td{u} = u /u_{max}$ and the black brane factor 
is in terms of $\td{u}$
\be
f (\td{u} ) = 1 - \ls \fr{\td{u}}{\td{u}_h} \rs^{\fr{c}{2 a_1 -1}} ,
\ee
with $\td{u}_h = u_h / u_{max}$ .
It should be noted that the above U-shape configuration is only possible when $u_{max}$ is
smaller than $u_h$, in other words, $\td{u}_h > 1$. If not, the turning point of the strip goes inside of the black brane
horizon. In this case, the resulting string configuration is described by two disconnected planes outside of the black brane horizon.

At zero temperature, $\td{u}_h \to \inf$ and $f(\td{u}) \to 1$ respectively. So the distance
$l$ and the area of the strip $A$ simply reduce to
\bea
l &=&   g_0  \ u_{max} , 	\la {res:zerotempdis}\\
A &=&  \fr{\td{h}_0}{ u_{max}^{D -3 - \th}},
\eea 
with
\bea
g_0 &=& \fr{2 \sqrt{\pi} \ \Gamma \ls \fr{ D - 1 -\th}{2 ( D - 2 -\th ) } \rs }{
 \Gamma \ls \fr{1}{2 ( D - 2 -\th ) } \rs}  ,\\
\td{h}_0 &=&  
-  \fr{2 L^{D-3}}{ (D -3 - \th) } \fr{1 }{ \e^{D - 3 - \th}  }  + h_0 ,  \la{res:entar}  \\
h_0 &=& \fr{2 L^{D-3}}{ (D -3 - \th) }
 \fr{\sqrt{\pi}  \Gamma \ls \fr{ D - 1 -\th}{2 ( D - 2 -\th ) } \rs }{ \G \ls  \fr{1}{2 ( D - 2 -\th )} \rs } ,
\eea
where the first term in \eq{res:entar} represents a UV divergence as $\e \to 0$. 
Ignoring the UV divergence, the zero temperature entanglement entropy
reduces to
\be
S_{en} =
 \fr{  L^{D-3} \lb 8 - (D-2) \et^2 \rb  }{2 G \lb 8 (D-3) + (D-2) \et^2 \rb} \ \ls \fr{2}{l} \rs^{ \fr{8 (D-3) + (D-2) \et^2}{8 - (D-2) \et^2}}
\lb \fr{\sqrt{\pi} \G \ls\fr{8 (D-1) - (D-2) \et^2}{16 (D-2)} \rs }{\G 
\ls \fr{8 - (D-2) \et^2}{16 (D-2)} \rs} \rb^{\fr{8 (D-2)}{8 - (D-2) \et^2} } .
\ee

If the size of the subsystem is small ($\td{u}_h >> 1$), after removing the
UV divergence, the distance and area of strip can be expanded into
\bea
l &=& u_{max}  \lb g_0 +  g_1   \
  \ls \fr{u_{max}}{u_h} \rs^{\zeta} \rb + \cdots , 
  \la{rel:disstrip}\\
A &=& \fr{1}{u_{max}^{\xi}} \lb h_0  + h_1  \
  \ls \fr{u_{max}}{u_h} \rs^{\zeta} \rb
   + \cdots , \la{rel:areastr}
\eea
where
\bea
\zeta &=& \fr{8 (D-1) - (D-2) \et^2}{8 - (D-2) \et^2} , \nn
\xi &=&  \fr{8 (D-3) +(D- 2) \et^2}{8 - (D-2) \et^2}, \nn
g_1 &=& \fr{\sqrt{\pi}  \lb 8 - (D-2) \et^2 \rb \ 
\G \ls \fr{8 (2 D-3) - (D-2) \et^2}{8 (D - 2)} 
\rs}{2  \lb 8(D-1) - (D-2) \et^2  \rb \ \G \ls \fr{4 (3 D-4)  - (D-2) \et^2}{8 (D - 2)} \rs}, \nn
h_1 &=& - \fr{L^{D-3} \sqrt{\pi} \G \ls \fr{8 (D-1)  - (D-2) \et^2}{8 (D-2)} \rs}{
 2 \G \ls \fr{ 4 D - (D-2) \et^2}{8 (D-2)} \rs }   .
\eea
In order to describe the entanglement entropy in terms of the system size, 
we need to rewrite $u_{max}$ in terms of $l$ and $u_h$. To do so,  we first 
set $u_{max}$ to  
\be		\la{rel:umaxl}
u_{max} = \fr{l}{g_0} (1 + \d) ,
\ee
where $\d$ is a small function of $l$ and $u_h$. Then, 
the first term satisfies the zero temperature result in \eq{res:zerotempdis}, whereas 
the second corresponds to the leading thermal correction.  
In order to satisfy \eq{rel:disstrip} at least at order of $u_h^{-{\zeta}}$, $\d$ 
should be
\be
\d = - \fr{g_1}{g_0^{{\zeta}+1}} \fr{l^{\zeta}}{u_h^{\zeta}} . 
\ee
When substituting this result into \eq{rel:areastr}, the area of strip becomes up to order of
$u_h^{-{\zeta}}$
\be
A = \fr{g_0^{\xi} \ h_0}{ l^{\xi}} \ls 1 + M \fr{l^{\zeta}}{u_h^{\zeta}} \rs ,
\ee
where $M$ is given by
\be
M = \fr{\xi \ g_1}{g_0^{{\zeta}+1}} + \fr{h_1}{h_0 \ g_0^{\zeta}} .
\ee
Using the following relation together with \eq{rel:umaxl}
\be
u_h  = \fr{8 (D-1)+ (D-2) \et^2}{4 \pi (8 +(D-2) \et^2)} \ 
\ls \fr{8 - (D-2) \et^2}{8 +(D-2) \et^2}  \rs^{\fr{8 - (D-2) \et^2}{(D-2) \et^2}}
\ \fr{1}{T_H} ,
\ee
the entanglement entropy in the small size limit leads to
\be			\la{res:enentropy}
S_{en} \equiv  \fr{A}{4 G} = \fr{g_0^{\xi} \ h_0}{4 G} \ls l^{-\xi} + \fr{M}{N^c} \ l^{\zeta-\xi} \ T_H^{\zeta} \rs ,
\ee
where the first term represents the entanglement entropy at zero temperature and the second is
the  leading thermal correction. In other words, the leading temperature-dependent entanglement entropy is proportional to 
\be
\D S_{en} \sim l^2 \ T_H^{\fr{8 (D-1) - (D-2)\et^2}{8 - (D-2)\et^2}} ,
\ee
which corresponds to the entropy increase caused by the excited states. 
Using the fact that the boundary energy density in \eq{res:holographicem} is proportional to $r_h^c$ only, we can easily  evaluate the total energy of strip when the strip distant $l$ in  \eq{ans:strip} is very small
\be
\D E = \int d^{D-2} x \  \fr{1}{8 \pi G}  \fr{4 (D-2)}{8 + (D-2) \et^2}  r_h^c
\sim \  l \ T_H^{\fr{8 (D-1) - (D-2)\et^2}{8 - (D-2)\et^2}} .
\ee 
These results show that the leading temperature-dependent entanglement entropy and
the energy of the excited states depend nontrivially on the Hawking temperature $T$,
which is the temperature of the thermal equilibrium. However, the ratio of them
is independent of the Hawking temperature 
\be
\fr{\D S_{en}}{\D E} \sim l ,
\ee
which shows the universal feature conjectured in \cite{Bhattacharya:2012mi}.
Introducing a entanglement temperature inversely proportional to the distance $l$,
the first law of thermodynamics is satisfied even in the relativistic non-conformal theory.


\section{Discussion}

In the $D$-dimensional Einstein-dilaton theory with a Liouville potential, 
a Schwarzschild-type black brane solutions is allowed and its asymptotic geometry has the 
$ISO(1,D-2)$ symmetry group. Following the gauge/gravity duality, this isometry group
can be reinterpreted as the Poincare group of the dual theory defined on the boundary.
In this case, because there is no scaling symmetry the dual theory becomes a relativistic 
non-conformal theory.  
In this paper, we have investigated the thermodynamic properties of the relativistic non-conformal
theory by using the holographic renormalization.
After introducing a correct counter term, we have evaluated the finite boundary stress tensor
and showed that the thermodynamics derived from it coincides with that of the EdBB geometry. Furthermore we showed, after identifying
the radial coordinate with the energy scale of the dual theory, that the free energy of the 
relativistic non-conformal theory monotonically decreases along the change of the energy from UV to IR.

For checking the self-consistency of the gauge/gravity duality in the Einstein-dilaton theory,
we studied the hydrodynamics of the dual theory by using the membrane paradigm.
The resulting transport coefficients coincide with those obtained by the Kubo formula. Furthermore, the thermodynamic quantity, $\e + P$, read off from the
momentum diffusion constant is also consistent with the QFT result. The charge and momentum diffusion constants monotonically decreases as the energy of the dual theory decreases.

Although the microscopic theory dual to the EdBB and the exact map between the bulk fluctuations and their dual operators are still unclear for the non-AdS geometry, our work shows 
the possibility to generalize the AdS/CFT correspondence to the non-AdS space. 
Based on such self-consistencies of the thermodynamic and macroscopic properties,
we further investigate some physical properties of the relativistic non-conformal theory,
the binding energies of particles and monopoles and the drag forces of them.
Due to the non-trivial coupling constant described by the dilaton, 
a particle and monopole have different physical properties in the non-conformal medium 
whereas they are indistinguishable in the conformal theory,
For example, for $\et=1$ the binding energy of monopoles is stronger than that of
particles. When the motion of a particle and monopole is non-relativistic
in the 4-dimensional relativistic non-conformal medium, the momentum of a particle
dissipates with a power law while a monopole has the dissipation with an inverse power law.
For a non-relativistic particle and monopole, the momentum dissipates exponentially.
In all cases, the dissipation rate is given by a function of the non-conformality.
We lastly showed that the universal feature of the entanglement temperature also
appears in the relativistic non-conformal theory.

The gauge/gravity duality in the non-AdS space is one of the important issues, so 
it remains interesting to investigate the microscopic aspects of it 
and to apply it to the real physical systems.
We hope to report more results in the future works.

\vspace{1cm}

{\bf Acknowledgement}

This work was supported by the Korea Science and Engineering Foundation
(KOSEF) grant funded by the Korea government(MEST) through the Center for
Quantum Spacetime(CQUeST) of Sogang University with grant number R11-2005-021.
C. Park was also
supported by Basic Science Research Program through the
National Research Foundation of Korea(NRF) funded by the Ministry of
Education, Science and Technology(2010-0022369).
\vspace{1cm}


\end{document}